# A Novel ground state in a new dimer iridate $Ba_{13}Ir_6O_{30}$ with $Ir^{6+}$ ($5d^3$) Ions

Hengdi Zhao[1], Feng Ye[2], Hao Zheng[1], Bing Hu[1,3], Yifei Ni[1], Yu Zhang[1],

Itamar Kimchi[1,4] and Gang Cao[1*]

[1] Department of Physics, University of Colorado, Boulder, CO 80309, USA

[2] Neutron Scattering Division, Oak Ridge National Laboratory, Oak Ridge, TN 37831, USA

[3] School of Mathematics and Physics, North China Electric Power University, Beijing 102206, China

[4] JILA, NIST and Department of Physics, University of Colorado, Boulder, CO 80309, USA

We have synthesized and studied a new iridate, $Ba_{13}Ir_6O_{30}$, with unusual Ir oxidation states: 2/3 $Ir^{6+}$($5d^3$) ions and 1/3 $Ir^{5+}$($5d^4$) ions. Its crystal structure features dimers of face-sharing $IrO_6$ octahedra, and $IrO_6$ monomers, that are linked via long, zigzag Ir-O-Ba-O-Ir pathways. Nevertheless, $Ba_{13}Ir_6O_{30}$ exhibits two transitions at $T_{N1}$ = 4.7 K and $T_{N2}$ = 1.6 K. This magnetic order is accompanied by a huge Sommerfeld coefficient 200 mJ/mole K below $T_{N2}$, signaling a coexisting frustrated/disordered state persisting down to at least 0.05 K. This iridate hosts unusually large $J_{eff}$=3/2 degrees of freedom, which is enabled by strong spin-orbit interactions (SOI) in the monomers with $Ir^{6+}$ ions and a joint effect of molecular orbitals and SOI in the dimers occupied by $Ir^{5+}$ and $Ir^{6+}$ ions. Features displayed by the magnetization and heat capacity suggest that the combination of covalency, SOI and large effective spins leads to highly frustrated ferrimagnetic ordering, possibly into a skyrmion crystal, a novelty of this new high-spin iridate.

*Email: gang.cao@colorado.edu

The inherently strong spin orbit interactions (SOI) in iridates create an entirely new hierarchy of energy scales and unique competitions between fundamental interactions (e.g., SOI ~ 0.4 eV, Coulomb interaction U ~ 0.5 eV) that result in intriguing consequences, such as the SOI-entangled $J_{eff}$ =1/2 state [1-8]. In essence, the strong SOI split the broad *5d* bands into two narrow bands, namely, $J_{eff}$=1/2 and $J_{eff}$ = 3/2 bands [1]. The $Ir^{4+}$(*5d$^5$*) ion provides four *d*-electrons to fill the lower $J_{eff}$ = 3/2 bands, and one electron to partially fill the upper $J_{eff}$ = 1/2 band that lies closest to the Fermi energy and dominates underlying physical properties. Indeed, the $J_{eff}$=1/2 state with spin S=1/2 and effective orbital moment $L_{eff}$ =1 is a building block essential to a panoply of novel phases observed or proposed in recent years [4-8]. The $Ir^{5+}$(*5d$^4$*) ion, on the other hand, provides all four *d*-electrons to fill the lower $J_{eff}$ = 3/2 bands, leaving the upper $J_{eff}$ = 1/2 band empty, thus S=1 and $L_{eff}$ =1 or a nonmagnetic singlet $J_{eff}$= 0 state. However, growing evidence indicates novel magnetic states emerging in pentavalent iridates when exchange interactions (0.05 - 0.10 eV)**,** singlet-triplet splitting (0.050 - 0.20 eV), non-cubic crystal fields and SOI are comparable and compete [9-15].

Nevertheless, studies of iridates have almost exclusively focused on tetravalent $Ir^{4+}$(*5d$^5$*) and pentavalent $Ir^{5+}$(*5d$^4$*) iridates thus far. There exist some pioneering studies on hexavalent $Ir^{6+}$(*5d$^3$*) iridates addressing structural and magnetic properties of the iridates [16-26]. However, our knowledge on the physics of hexavalent iridates is conspicuously lacking, in part, because such a high oxidation state is not as common as $Ir^{4+}$ and $Ir^{5+}$ in iridates. A *d$^3$* ion in an octahedron normally generates an S=3/2 state (with the quenched orbital moment L=0) when the Hund's ruling coupling $J_H$ is much stronger than SOI. However, for the $Ir^{6+}$(*5d$^3$*) ion, the SOI is almost as strong as that in the $Ir^{4+}$(*5d$^5$*) and $Ir^{5+}$ ($5d^4$) ions [25]; accordingly, three $d$-electrons in the $Ir^{6+}$ ion fill three of the four orbitals determined by the *j-j* coupling, forming a SOI-entangled $J_{eff}=3/2$ state. The $J_{eff}=3/2$ state is recently observed in $Os^{5+}$ ($5d^3$) based oxides [26]. Furthermore, the overwhelming balance of interest has been given to iridates with either corner-sharing octahedra (e.g., perovskite iridates) or edge-sharing octahedra (e.g., honeycomb iridates). Very limited work on iridates with facing-sharing octahedra has been conducted until very recently [27-32]. It needs to be emphasized that the Ir-Ir bond distance in facing-sharing octahedra ($2/\sqrt{3}$ $d_{Ir-O}$ $\approx 1.16 d_{Ir-O}$, where $d_{Ir-O}$ is the Ir-O bond distance in an undistorted octahedron) is shorter than that both in corner-sharing octahedra ($2d_{Ir-O}$) and edge-sharing octahedra ($\sqrt{2}$ $d_{Ir-O}$). There is mounting evidence that suggests that the shorter Ir-Ir bond distance facilitates a strong covalency/electron hopping, which, along with the SOI, may significantly alter or diminish the $J_{eff}$ states, resulting in new quantum phases in dimer systems [27-32].

Here we report an exotic magnetic state in a newly synthesized single-crystal iridate $Ba_{13}Ir_6O_{30}$. This compound, which did not exist previously, consists of 2/3 hexavalent $Ir^{6+}$ ($5d^3$) ions and 1/3 pentavalent $Ir^{5+}$ ($5d^4$) ions, rendering an average oxidation state of $Ir^{+5.67}$. The crystal structure of the compound features two slightly different dimers of face-sharing $IrO_6$ octahedra and additional $IrO_6$ monomers. There is no obviously strong connectivity that links the dimers and/or monomers but long, zigzag Ir-O-Ba-O-Ir pathways (**Fig.1a**). However, $Ba_{13}Ir_6O_{30}$ strikingly exhibits two antiferromagnetic (AFM) transitions at $T_{N1} = 4.7$ K and $T_{N2} = 1.6$ K, respectively, indicating established three-dimensional correlations. Hysteresis at $T_{N1}$ suggests a possible ferrimagnetic order or magnetic canting. The average effective moment $\mu_{eff}$ is ~ 4.50 $\mu_B$/f.u. which, as we discuss below, requires at least some of the local moments to be larger than spin-one-half, consistent with Hund's rule coupling ($J_H \sim 0.7$ eV) [28] playing a role constrained by SOI. Interestingly, the $T_{N1}$ ordered moment $\mu_s$ is 1.14 $\mu_B$/f.u. and the entropy removal is only ~ 3.7 J/mole K below 5 K; below $T_{N2}$, a huge Sommerfeld coefficient $\gamma = 200$ mJ/mole $K^2$ is observed, signaling gapless excitations suggestive of a frustrated/disordered phase persisting down to at least 0.05 K. This behavior is attributed to an incomplete removal of the large spin degrees of freedom unique to $Ba_{13}Ir_6O_{30}$. We argue that an unusual spin-orbit-entangled $J_{eff}=3/2$ state is hosted by the monomers with $Ir^{6+}$ ions or monomers$^{6+}$, while a distinct quasi-spin state dictated by molecular orbitals is hosted by the dimers with both $Ir^{6+}$ and $Ir^{5+}$ ions. This work, along with comparisons drawn with other dimer iridates, suggests that a combined effect of the strong covalency and SOI suppresses conventional $J_{eff}$ states, leading to the exotic quantum ground state with large residual entropy.

$Ba_{13}Ir_6O_{30}$ adopts an orthorhombic structure with a Pbcn (No. 60) space group. The lattice parameters at 100K are $a = 25.4716(7)$, $b = 11.7548(3)$ Å and $c = 20.5668(6)$ Å, respectively. The unit cell consists of 48 Ir ions (**Fig.1**) (see Supplemental Material (SM) for more details [33]). Since the structure of this compound was never reported before, an intensive structural study of the single-crystal $Ba_{13}Ir_6O_{30}$ at different temperatures has been conducted using both single-crystal X-ray diffraction and neutron diffraction to ensure an accurate determination of the crystal structure of this new compound [33]. The single crystals of $Ba_{13}Ir_6O_3$ are clearly of high quality, as shown in **Figs.1d** and **1e**.

The first striking feature of $Ba_{13}Ir_6O_{30}$ involves its unusual magnetic oxidation states. $Ba_{13}Ir_6O_{30}$ hosts 2/3 hexavalent $Ir^{6+}$ (*5d³*) ions and 1/3 pentavalent $Ir^{5+}$(*5d⁴*) ions which give rise to an average $Ir^{+5.67}$ oxidation state. There are two slightly different dimers. The Ir1 and Ir3 ions form the shorter dimer or Dimer S and the Ir2 and Ir4 ions form the longer one or Dimer L (**Fig. 1b**). At 100 K, the Ir1-Ir3 bond distance of the Dimer S (=2.752 Å) is shorter than the Ir2-Ir4 bond distance of the Dimer L (=2.789 Å) by 1.3%. Both Dimers L and Dimer S are alternatively arranged along the *c* axis (**Fig.1b**). Each dimer consists of two slightly inequivalent octahedral sites. The $IrO_6$ monomers are highly distorted but their average Ir-O bond distance fluctuates around 1.92 Å, expected for $Ir^{6+}$ ions [16]. Thus, all monomers host a hexavalent $Ir^{6+}$($5d^3$) ion. Each $IrO_6$ octahedron has a strong trigonal distortion in the dimers. There are eight Dimer S's, eight Dimer L's and sixteen $IrO_6$ monomers in a unit cell [33]. It is emphasized that the dimers and monomers are only linked by long, zigzag Ir-O-Ba-O-Ir pathways in the *bc* plane (**Figs.1a** and **1c**).

Despite the unfavorable connectivity for three dimensional correlations, $Ba_{13}Ir_6O_{30}$ undergoes two magnetic transitions below 5 K. The magnetization M for three principal axes exhibits an abrupt rise at $T_{N1}$ = 4.7 K and anomalous behavior leading to another transition at $T_{N2}$ = 1.6 K (see **Fig.2a**, and heat capacity C(T) discussed below). The observed large magnetic anisotropy indicates that the SOI is significant in spite of the trigonal distortion in the $IrO_6$ octahedra. The *c*-axis magnetization $M_c$ is also measured at various fields and exhibits a rounded but still well-defined anomaly near $T_{N1}$ at 7 T (**Fig.2a Inset**).

Data fits to the Curie-Weiss law yield a Curie-Weiss temperature $\theta_{CW}$ = - 63, - 40, and -27 K for the *a*, *b*, and *c* axis, respectively (**Fig.2b**). The negative $\theta_{CW}$ conforms to an AFM exchange interaction in the system. The corresponding frustration parameter, defined as *f* = $|\theta_{CW}|/T_N$, is as large as 13, signaling sizable frustration in the system. The effective moment $\mu_{eff}$ is 4.74, 4.29 and 4.56 $\mu_B$/f.u. for the *a*, *b*, and *c* axis, respectively. Given six Ir ions (two $Ir^{5+}$ ions and four $Ir^{6+}$ ions) per formula, these values are inconsistent with those expected for either the Hund's rule physics or the single-site physics of ideal isolated octahedra in which $J_{eff} = 0$ for $Ir^{5+}(5d^4)$ ions and $J_{eff} = 3/2$ for $Ir^{6+}(5d^3)$ ions. Instead, the two AFM transitions and the reduced $\mu_{eff}$ imply an existence of strongly coupled Ir pairs or molecular orbitals in this system, further discussed below.

The isothermal magnetization M(H) at 1.8 K exhibits a curvature with increasing H, suggesting a weak ferromagnetic behavior due to magnetic canting or the Dzyaloshinskii–Moriya (DM) interactions below $T_{N1}$ (**Fig.2c**). This is consistent with a strong hysteresis effect observed below $T_{N1}$ when the sample is measured in a zero-field cooled (ZFC) and a field cooled (FC) sequence, respectively (see **Fig.2c Inset**). Extrapolating $M_c$ to H=0 yields an ordered moment of 1.14 $\mu_B$/f.u.

The specific heat C(T) measured down to 0.05 K confirms the existence of the long-range magnetic orders. C(T) exhibits an anomaly at $T_{N1}$ = 4.7 K and a broader one at $T_{N2}$ = 1.6 K (see **Fig.3a**). The anomalies are better illustrated in a C/T vs T plot shown in **Fig.3b**. C(T) measured at $\mu_oH$ = 1 T shows a discernible increase in both transitions $T_{N1}$ and $T_{N2}$, and a broadening of $T_{N1}$. It is important to be pointed out that both $T_{N1}$ and $T_{N2}$ do not show a λ-transition expected for a typical $2^{nd}$-order phase transition such as the one observed in a related dimer iridate, $Ba_5AlIr_2O_{11}$ [27]. An analysis of the C(T) data yields an entropy removal of approximately 3.70 J/mole K below $T_{N1}$, significantly smaller than that anticipated for any spin systems, such as 5.76 J/mole K for an S = 1/2 system. Indeed, the Sommerfeld coefficient or γ extrapolated above $T_{N1}$ and below $T_{N2}$ are 480 and 200 mJ/mole K$^2$, respectively (see **Fig.3c**). Such surprisingly huge values of γ, which are conventionally expected in a highly correlated metal, signal a large amount of remaining entropy or an existence of a significantly frustrated/disordered state that persists down to at least 0.05 K coexisting with the ordered magnetic state.

The *b*-axis electrical resistivity $\rho_b$ between 100 K and 800 K exhibits an activation behavior with an activation charge-gap of 0.23 eV (**Fig.4**). $\rho_b$ becomes too high for measurements below 100 K and remains an excellent electrical insulator. $Ba_{13}Ir_6O_{30}$ shows no sign of charge order over this wide temperature range, in contrast to $Ba_5AlIr_2O_{11}$ in which a charge order occurs at 210 K [27].

It is clear that three-dimensional magnetic correlations are established at low temperatures, which means that all $IrO_6$ monomers and dimers must participate in the long-range magnetic order, despite the remarkably weak connectivity between them (**Fig.1a**). Recent studies have already established that the strong covalency coupled with large SOI within the dimers can be comparable or even stronger than the intraatomic parameters such as the Hund's rule coupling $J_H$, forming bonding and antibonding states or molecular orbitals [27-32]. Both Dimer S and Dimer L are apparently the main building blocks of the magnetic state.

$Ba_{13}Ir_6O_{30}$ shares some structural similarities to the dimer-chain system $Ba_5AlIr_2O_{11}$ [27]. The latter iridate features both tetravalent $Ir^{4+}$ and pentavalent $Ir^{5+}$ ions in each of dimers ($Ir^{4+}$-$Ir^{5+}$dimer) that are linked via $AlO_4$-tetrahedra. Despite their one-dimensional character, the dimer-chains of $Ba_5AlIr_2O_{11}$ undergo a magnetic order at $T_N$ = 4.5 K, remarkably close to $T_{N1}$ = 4.7 K, and a charge order at $T_s$ = 210 K facilitated by two significantly inequivalent Ir sites [27]. It is found that the magnetic moments are significantly suppressed because of the strong covalency and SOI, which lead to the formation of molecular orbitals [27, 28].

This comparison suggests that $T_{N1}$ and the reduced magnetic moments in $Ba_{13}Ir_6O_{30}$ may share the same origin as those in $Ba_5AlIr_2O_{11}$, which, in essence, is a result of the molecular orbitals in the dimers [28]. However, $Ba_{13}Ir_6O_{30}$ exhibits no sign of charge order. Without charge order, an even stronger covalency within each dimer is anticipated, and it favors a fractional occupation of *5d*-orbitals in a bonding state. In $Ba_{13}Ir_6O_{30}$, there are six orbitals consisting of two sets of *$t_{2g}$* orbitals with a different symmetry in each dimer. Each set of *$t_{2g}$* orbitals is split by the strong trigonal distortion into a lower *$a_1$* orbital and two higher *e* orbitals. The spatial orientation of these orbitals is such that the *$a_1$* orbitals form a lower-lying bonding combination and a corresponding higher-lying antibonding orbital *$a_1$*\* in each dimer. The four *e* orbitals reside between *$a_1$* and *$a_1$*\*. The antibonding orbital *$a_1$*\* is much more energetically costly, in part because the strong SOI pushes it further higher in energy [28]. The splitting between the bonding and antibonding orbitals is thus large, leaving the antibonding orbital *$a_1$*\* unoccupied (**Figs.5b** and **c**).

The oxidation states in the short and long dimers can be inferred from the associated interatomic distances. The Ir-Ir bond distance in Dimer L is only slightly longer than that of Dimer S by 1.3%. A larger difference (4 - 5%) would suggest [24] that the dimers differ in their oxidation state, for example, that Dimer S would host two $Ir^{5+}$ ions and the Dimer L two $Ir^{6+}$ ions (both likely resulting in an effective spin-1 state). While this scenario cannot be ruled out, a likelier scenario is that both Dimers S and L accommodate a mixed oxidation state of $Ir^{5+}$ and $Ir^{6+}$, thus dimers[11+]. The strong covalency favors a fractional occupation of *5d*-orbitals in a bonding state especially given the strong electron hopping in the dimers [27-30]. The mixed valent state is likely primarily dynamic rather than static since the difference between the two Ir sites in each dimer is very small [33]. A small overlap with the static 5+/6+ configuration could generate electric dipoles in the dimers, which could then alternate the direction in order to minimize the electrical repulsion, as shown in **Fig.1b**. This mostly-dynamic mixed valence scenario is consistent with the structural data [33].

There are multiple possible spin states in the dimers[11+] and the monomers[6+]. For the monomer, the SOI-entangled $J_{eff}$=3/2 state is expected [34] (**Fig.5a**), unless beyond-octahedral distortions become sufficiently strong to quench the spin to a $J_{eff}$=1/2 state . For the dimers, notice that each of them hosts seven *5d*-electrons. The likely possibility is that six of them will doubly occupy the lowest three orbitals and the remaining one electron will singly occupy the next higher orbital, giving rise to a quasi-spin S = 1/2 state (**Fig.5b**). A higher spin S=3/2 state is also possible (**Fig.5c**); given the observed $\mu_{eff}$, having an effective spin 3/2 state on both dimers and monomers requires an average g-factor that is reduced by strong SOI to about g=1.15, which is well within the range of observed g-values for ions with high atomic numbers [35]. Therefore, the quasi-spin 3/2 state for the dimers is a plausible scenario (**Fig.5c**). It is emphasized that an effective spin 1/2 state for both the monomer and dime is inconsistent with the observed $\mu_{eff}$, which is too large for this low-spin scenario, even assuming that the g-factors are as large as the bare spin value of g=2. We therefore conclude that at least one site – the monomer, or the dimer, or both – shows a large effective spin 3/2 state.

The large-spin state(s) across two distinct magnetic sites produces a large number of degrees of freedom (2S+1 each) which should explain the striking features, namely, (1) ferrimagnetism (**Fig.2c**), (2) two magnetic transitions (**Figs.2a** and **3a**), and (3) huge low-temperature γ=200 mJ/mole K² or entropy (**Fig.3c**).  A ferrimagnetic order is consistent with the negative (antiferromagnetic) Curie-Weiss temperatures (**Fig.2b inset**) while the hysteresis seen below $T_{N1}$ (**Fig.2c inset**) is consistent with some net ferromagnetic moment. Ferrimagnetism is generically expected due to the presence of two distinct magnetic sites, and in this case, the monomers and dimers. The two transitions and the large residual entropy that remains even below $T_{N2}$ suggest that the ordering patterns are incomplete. This is consistent with the large number of spin degrees of freedom present in this compound, where at least one of two sites hosts an effective spin 3/2 state.

Large-spin iridates with the complex interplay of correlations and SOI have been largely unexplored, and there is insufficient information to determine the magnetic order here. We can however point out a few possibilities that are consistent with the available experimental data: (A) ferrimagnetic order with different moments on dimers and monomers that undergoes a second transition with noncollinear canting; or (B) a skyrmion crystal. Skyrmion crystals [36] are known to be able to arise from various related mechanisms based on SOI, including **(i)** competition between single-ion spin anisotropy of large spin (greater than spin half) and antisymmetric DM interactions; and **(ii)** magnetic frustration in large-spin magnets with strong SOI [37, 38]. Both mechanisms suit the present setting of the spin 3/2 state arising from strong SOI: Mechanism **(i)** requires DM interactions which is allowed due to the absence of inversion symmetry on lines connecting dimers, associated with the low symmetry configuration of octahedral dimers; while the magnetic frustration of Mechanism **(ii)** arises naturally from the lattice formed by the multiple magnetic sites (monomers and short and long dimers). Interestingly, Mechanism **(ii)**, which is likely the primary mechanism, would result in a dense skyrmion lattice, with the skyrmion-lattice-scale related to the microscopic crystal scale by a factor of order one.

A skyrmion lattice is an appealing theoretical scenario for explaining the measurements since it would intrinsically show a net magnetic moment despite the primarily antiferromagnetic correlations, consistent with the hysteretic higher transition $T_{N1}$, and can easily produce a second ordering instability at a lower temperature $T_{N2}$. In addition, the large residual heat capacity indicates an order that is incomplete (partially quantum-disordered) and/or that hosts numerous low-energy excitations. Gapless phonons and other emergent crystal modes in a skyrmion crystal may offer a route to at least some of this residual entropy.

The large $\gamma$ remains the most surprising and intriguing experimental feature of this compound. Indeed, the value of $\gamma$ is nearly zero for the dimer-chain $Ba_5AlIr_2O_{11}$ that features a quasi-spin S=1/2 state [28] although it similarly orders at $T_N$ = 4.5 K [27]. A full characterization of the lowest temperature magnetic order may help shed light on this striking feature. We expect that the key ingredient for the large quantum entropy at low temperatures, as for the other striking experimental observations here, relies on the unusually large spin degrees of freedom unique to the structure of this iridate and, in particular, its monomers and dimers of $Ir^{6+}$ ($5d^3$) ions. All in all, this work provides a new paradigm for discovery and study of novel quantum states in spin-orbit-entangled, high-spin materials, which have remained largely unexplored.

*Acknowledgements* This work was supported by the National Science Foundation via Grants Nos. DMR-1712101. I.K. was supported by a National Research Council Fellowship through the National Institute of Standards and Technology.

**Captions:**

**Fig.1.** Crystal structure: **(a)** The dimers and monomers that are connected via long, zigzag Ir-O-Ba-O-Ir pathways along the $c$ axis. **(b)** The arrangement of dimers. Note that Ir1-Ir3 and Ir2-Ir4 are the Dimer S and Dimer L, respectively. **(c)** The bc plane to further illustrate the weak connectivity of the dimers and monomers. **(d)** A image of a representative single crystal of $Ba_{13}Ir_6O_{30}$ with the $bc$ plane. **(e)** Three representative images of the reciprocal lattice, which unambiguously reveal the high-quality of the single crystal and the space group (#60).

**Fig. 2.** Magnetic properties: Temperature dependence of **(a)** the magnetization M for the a-, b- and c-axis, $M_a$, $M_b$ and $M_c$ at $\mu_oH = 0.005$ T. The red dashed line for $M_c$ is guide for the eye. Inset: $M_c$ at various fields; **(b)** the magnetic susceptibility for the a-, b- and c-axis, $\chi_a$, $\chi_b$, and $\chi_c$ at $\mu_oH = 0.2$ T. Inset: the corresponding $\Delta\chi$ vs T. **(c)** The isothermal magnetization for the a-, b- and c-axis at 1.8 K. Inset: Mc vs T at $\mu_oH = 0.01$T.

**Fig. 3.** Heat capacity: Temperature dependence of **(a)** the specific heat C(T) for a temperature range of 0.05 – 10 K. Mc in red (right scale) is for comparison. The two gray dashed lines are marks for $T_{N1}$ and $T_{N2}$; **(b)** C/T at $\mu_oH=0$ and 1 T. Note the changes due to the applied field. **(c)** C/T vs $T^2$ below 0.6 K; Inset: C/T vs $T^2$ for 0.05 < T < 12 K. Note the large interception of the red dashed line on the vertical axis.

**Fig.4.** Temperature dependence of the b-axis electrical resistivity $\rho_b$. Inset: a fit to the activation law.

**Fig. 5.** A sketch for the proposed configurations of orbitals, electrons and total quasi-spins for **(a)** the $IrO_6$ monomer, **(b)** the $Ir^{5+}/Ir^{6+}$ Dimer with S=1/2, and **(c)** the $Ir^{5+}/Ir^{6+}$ Dimer with S=3/2.

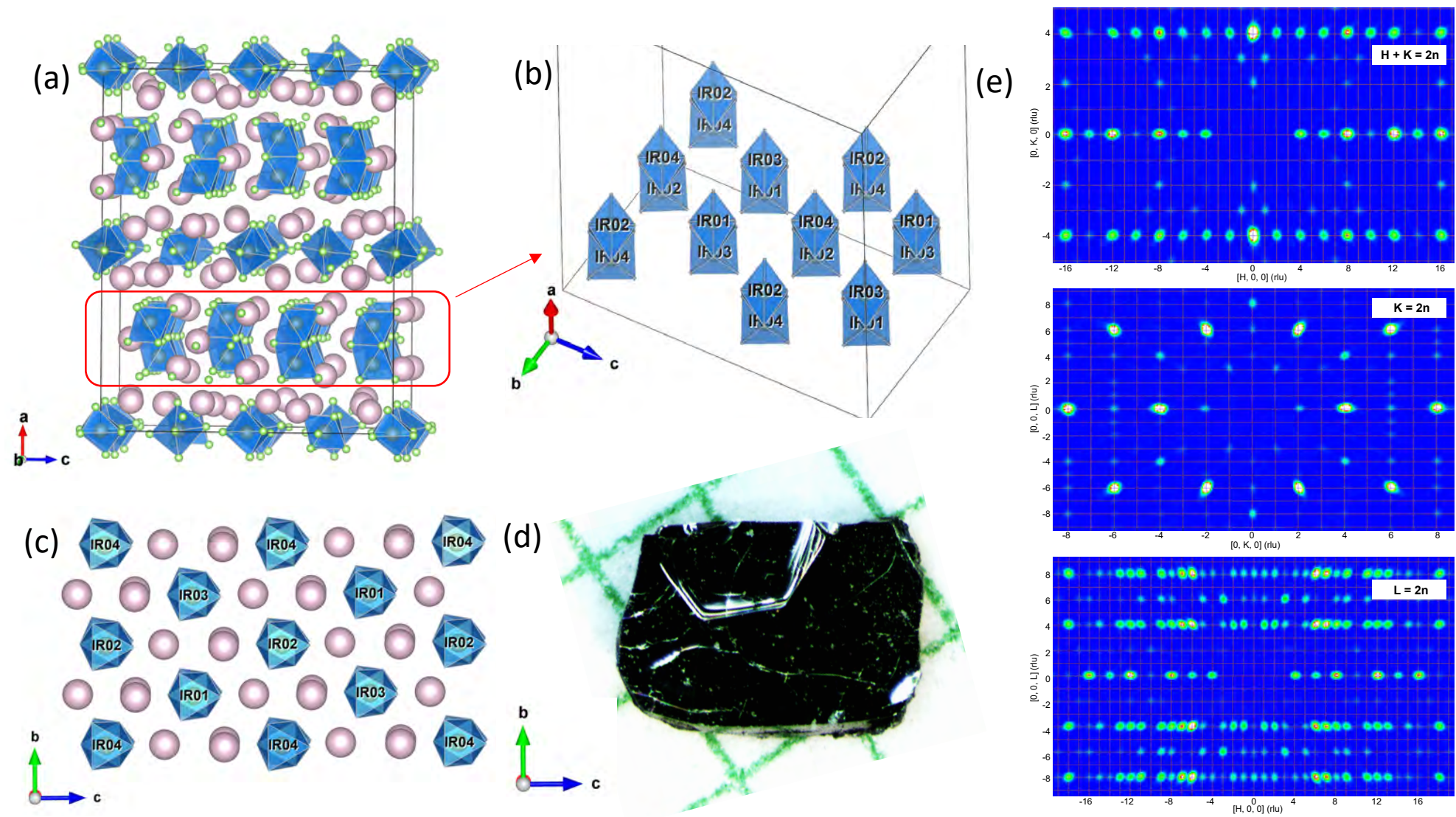

Fig. 1

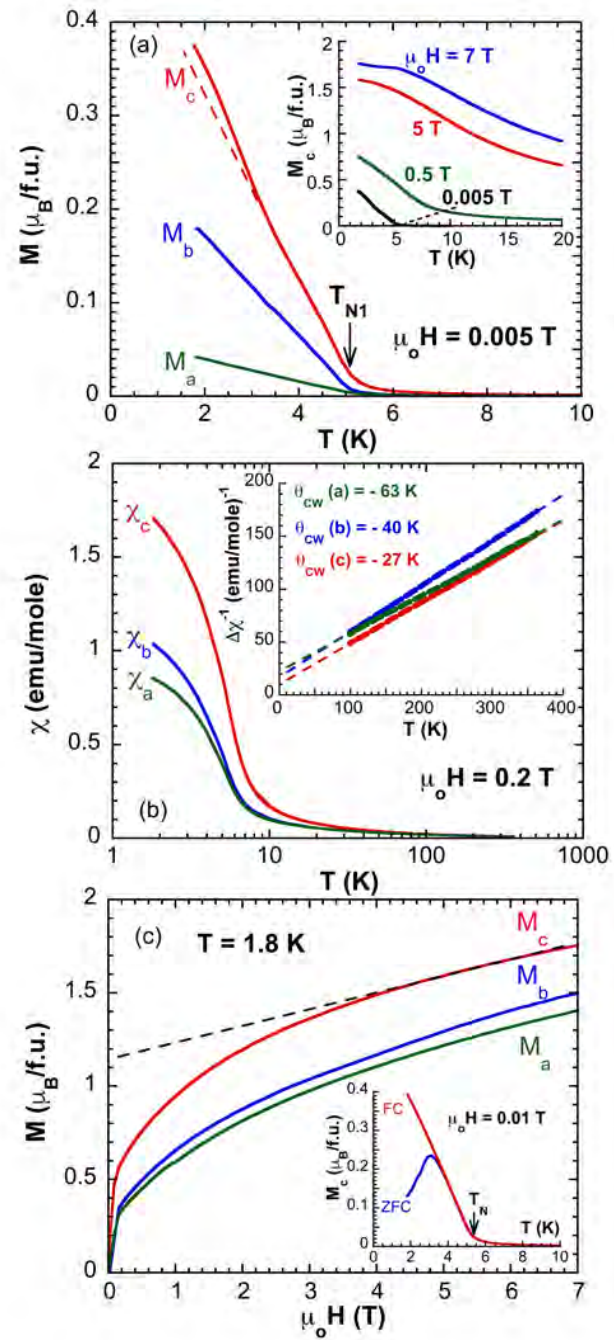

Fig. 2

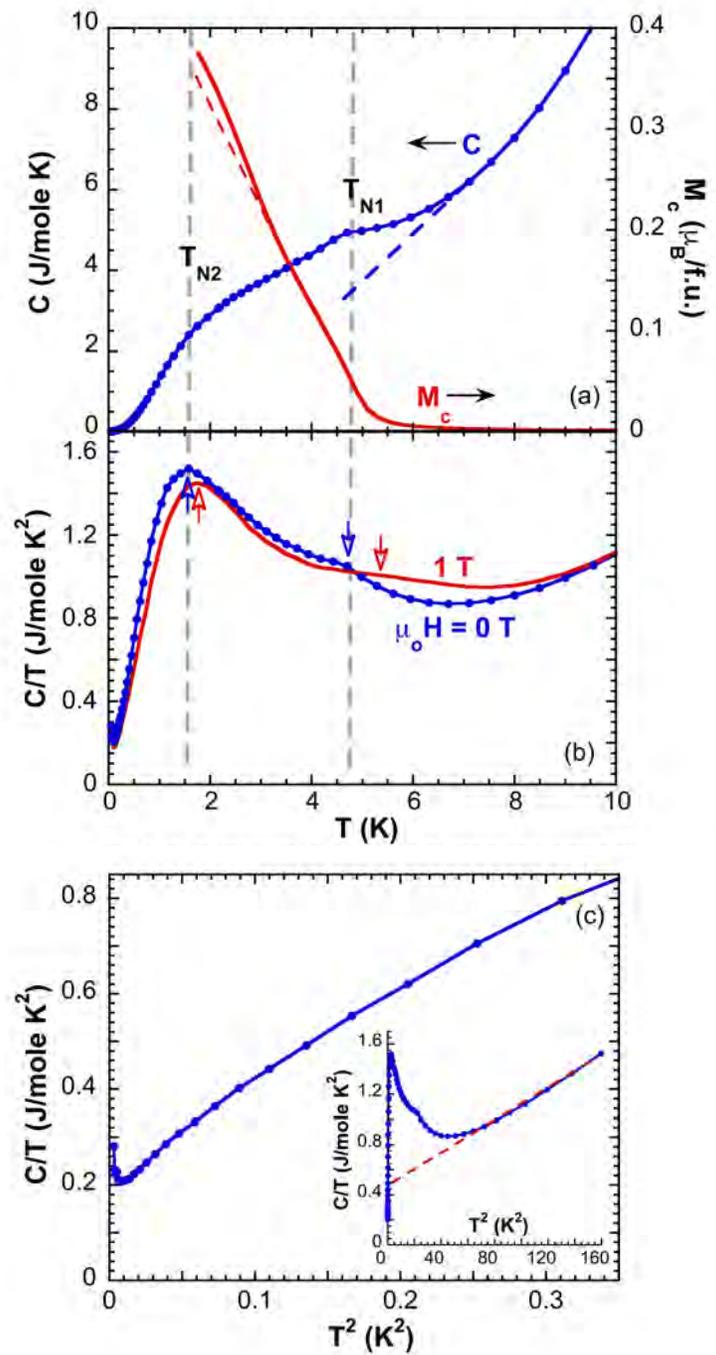

Fig. 3

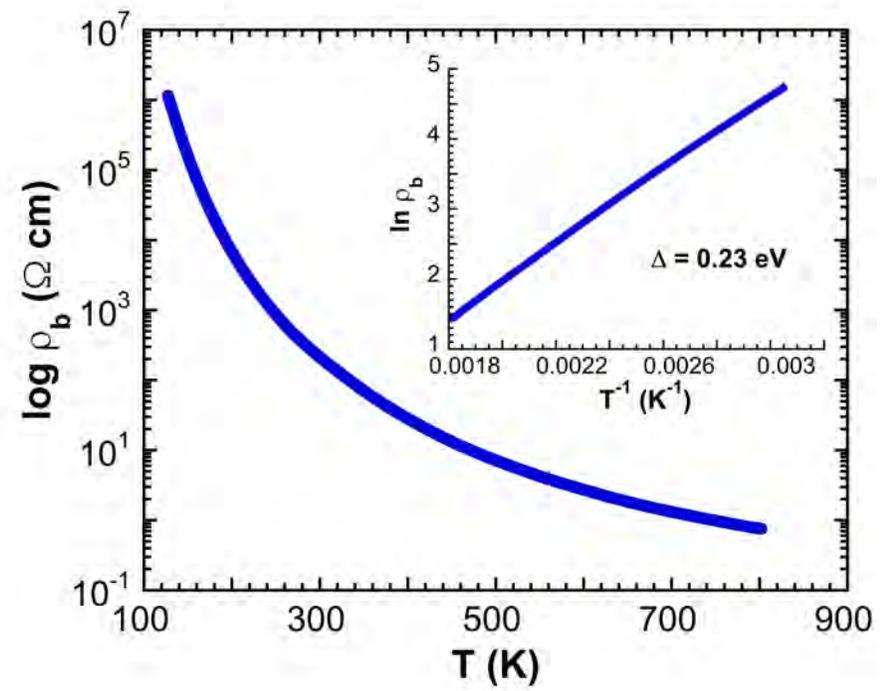

Fig. 4

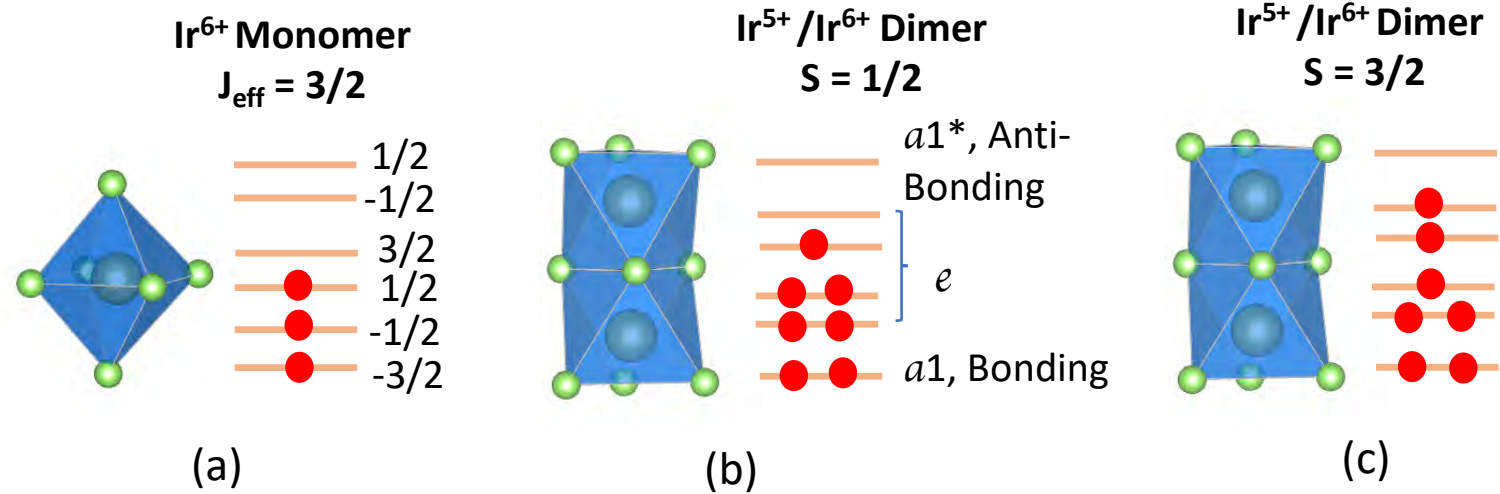

Fig. 5

# A Novel ground state in a new dimer iridate $Ba_{13}Ir_6O_{30}$ with $Ir^{6+}(5d^3)$ Ions

Hengdi Zhao[1], Feng Ye[2], Hao Zheng[1], Bing Hu[1,3], Yifei Ni[1], Yu Zhang[1],

Itamar Kimchi[1,4] and Gang Cao[1]

[1] Department of Physics, University of Colorado, Boulder, CO 80309, USA

[2] Neutron Scattering Division, Oak Ridge National Laboratory, Oak Ridge, TN 37831, USA

[3] School of Mathematics and Physics, North China Electric Power University, Beijing 102206, China

[4] JILA, NIST and Department of Physics, University of Colorado, Boulder, CO 80309, USA

## *Experimental Details*

Single crystals of $Ba_{13}Ir_6O_{30}$ were grown using a flux method from off-stoichiometric quantities of $IrO_2$, $BaCO_3$, and $BaCl_2$. The average sample size is approximately 3 x 2 x 1 $mm^3$. Measurements of crystal structures were performed using a Bruker Quest ECO single-crystal diffractometer equipped with a PHOTON 50 CMOS detector. It is also equipped with an Oxford Cryosystem that creates sample temperature environments ranging from 80 K to 400 K during x-ray diffraction measurements. All datasets were reduced and refined by using APEX3 and SHELX-2014 program [1]. Chemical analyses of the samples were performed using a combination of a Hitachi MT3030 Plus Scanning Electron Microscope and an Oxford Energy Dispersive X-Ray Spectroscopy (EDX). Magnetic properties were measured using a Quantum Design (QD) MPMS-7 SQUID Magnetometer. Standard four-lead measurements of the electrical resistivity were carried out using a QD Dynacool PPMS System equipped with a 14-Tesla magnet. The heat capacity was measured down to 0.05 K using a dilution refrigerator for the PPMS.

The crystal structure of $Ba_{13}Ir_6O_{30}$ was also studied using single-crystal neutron diffraction using the time-of-flight Laue diffractometer TOPAZ. The data were collected on crystal with a volume of approximately 1.0 mm$^3$ for 70 minutes at every orientation. A total of approximately 20 hours was spent to collect the data at 100 K. Sample orientations were optimized for an estimated 99% coverage of symmetry-independent reflections of the monoclinic cell. The raw Bragg intensities were obtained using a 3D ellipsoidal integration method. Data reduction including Lorentz, absorption, TOF spectrum, and detector efficiency corrections were carried out. The reduced data were exported to the GSAS program suite for wavelength dependent extinction correction and refined to convergence using SHELXL-97.

*Crystal Structure*

Ba$_{13}$Ir$_6$O$_{30}$ adopts an orthorhombic structure with a Pbcn (No. 60) space group or a pseudo-hexagonal structure. The lattice parameters at 100K are $a$ = 25.4716(7) $b$ = 11.7548(3) Å, $c$ = 20.5668(6) Å (see **Table I** for details).

Ba$_{13}$Ir$_6$O$_{30}$ consists of two pairs of face-sharing IrO$_6$ octahedral dimers and four single IrO$_6$ monomers in each unit cell. It hosts 2/3 Ir$^{6+}$(*5d$^3$*) ions and 1/3 Ir$^{5+}$ (*5d$^4$*) ions. One distinctive feature of this compound is that all the dimers and monomers seem isolated with each other, with only Ba ions in between (see **Fig.1**). A complete set of the Ir-O bond distances inside dimers and monomers is tabulated in **Table II** and **Table III**, respectively.

**References**
1. G. M. Sheldrick, Acta Cryst A **64**, 112 (2008)

**Table I.** Structural parameters for single crystal Ba$_{13}$Ir$_6$O$_{30}$ in *Pbcn* (No. 60) phase at 100 K, data obtained from single crystal X-ray diffraction.

| T = 100 K | $a$ = 25.4716(7) $b$ = 11.7548(3) Å, $c$ = 20.5668(6) Å, Z=8, V = 6158.0(5) Å$^3$. The agreement factor R$_1$ = 6.19% was achieved by using 16262 unique reflections with I > 4σ and resolution of $d_{min}$ =0.55 Å. Anisotropic atomic displacement parameters were used for all elements except O15 and O155. | | | | | |
|---|---|---|---|---|---|---|
| | x | y | z | Occupancy | Ueq(Å$^2$) | site |
| Ir01 | 0.31258(2) | 0.25177(3) | 0.91408(2) | 1 | 0.00943(5) | 8d |
| Ir02 | 0.19847(2) | 0.49989(3) | 0.66602(2) | 1 | 0.00945(5) | 8d |
| Ir03 | 0.29546(2) | 0.24626(3) | 0.41453(2) | 1 | 0.00926(5) | 8d |
| Ir04 | 0.30795(2) | 0.49646(3) | 0.66455(2) | 1 | 0.00960(5) | 8d |
| Ir05 | 0.5 | 0 | 0.5 | 1 | 0.01226(8) | 4b |
| Ir06 | 0.5 | 0.24136(4) | 0.75 | 1 | 0.01159(7) | 4c |
| Ir07 | 0.5 | -0.25964(5) | 0.75 | 1 | 0.01378(8) | 4c |
| Ir08 | 0.5 | 0.5 | 1 | 1 | 0.01595(9) | 4a |
| Ba09 | 0.17294(2) | 0.2513(4) | 0.74837(2) | 1 | 0.01011(8) | 8d |
| Ba0A | 0.25003(2) | 0.50208(4) | 0.82965(2) | 1 | 0.01054(8) | 8d |
| Ba0B | 0.32816(2) | -0.0017(4) | 0.49974(2) | 1 | 0.01101(8) | 8d |
| Ba0C | 0.25055(2) | 0.24791(4) | 0.58447(3) | 1 | 0.01089(8) | 8d |
| Ba0D | 0.41997(2) | 0.54212(6) | 0.84483(3) | 1 | 0.0182(1) | 8d |
| Ba0E | 0.16699(2) | -0.00491(5) | 0.49974(3) | 1 | 0.01372(9) | 8d |
| Ba0F | 0.34443(3) | 0.25086(5) | 0.74901(3) | 1 | 0.0156(1) | 8d |
| Ba0G | 0.55632(3) | 0.22768(7) | 0.5978(3) | 1 | 0.0208(1) | 8d |
| Ba0H | 0.43908(3) | 0.46946(7) | 0.67198(4) | 1 | 0.0239(1) | 8d |
| Ba0I | 0.42093(3) | 0.75948(6) | 0.58652(4) | 1 | 0.0233(1) | 8d |
| Ba0J | 0.41868(3) | 0.23299(8) | 0.55478(4) | 1 | 0.0264(2) | 8d |
| Ba0K | 0.43215(3) | 0.01614(8) | 0.68746(5) | 1 | 0.0292(2) | 8d |
| Ba0L | 0.42973(5) | -0.0768(1) | 0.85923(6) | 1 | 0.0519(4) | 8d |
| O8 | 0.2587(3) | 0.2540(6) | 0.8394(3) | 1 | 0.0113(9) | 8d |
| O7 | 0.2428(3) | 0.3589(6) | 0.4520(3) | 1 | 0.012(1) | 8d |
| O6 | 0.2527(3) | 0.4974(5) | 0.5911(3) | 1 | 0.0112(9) | 8d |
| O151 | 0.3537(3) | 0.3699(5) | 0.8768(4) | 1 | 0.015(1) | 8d |
| O152 | 0.4580(4) | 0.1232(8) | 0.79123(4) | 1 | 0.023(2) | 8d |
| O153 | 0.4978(6) | 0.416(1) | 0.9235(7) | 1 | 0.046(3) | 8d |
| O154 | 0.5760(6) | -0.256(2) | 0.753(2) | 1 | 0.10(1) | 8d |
| O155 | 0.5 | -0.107(4) | 0.75 | 1 | 0.11(1) | 4c |
| O141 | 0.3479(3) | 0.3758(6) | 0.6247(4) | 1 | 0.014(1) | 8d |
| O142 | 0.3506(3) | 0.6114(7) | 0.6242(4) | 1 | 0.016(1) | 8d |
| O146 | 0.4872(5) | 0.037(1) | 0.5877(5) | 1 | 0.034(2) | 8d |
| O131 | 0.1560(3) | 0.3810(6) | 0.6287(3) | 1 | 0.013(1) | 8d |
| O13 | 0.3531(3) | 0.2504(7) | 0.9917(4) | 1 | 0.016(1) | 8d |
| O112 | 0.1571(3) | 0.6173(6) | 0.6267(4) | 1 | 0.014(1) | 8d |
| O12 | 0.3495(3) | 0.1313(6) | 0.8744(3) | 1 | 0.014(1) | 8d |
| O111 | 0.2411(3) | 0.1348(6) | 0.4519(3) | 1 | 0.013(1) | 8d |
| O11 | 0.2531(3) | 0.6120(6) | 0.7015(3) | 1 | 0.012(1) | 8d |
| O10   | 0.1607(3) | 0.5012(6)  | 0.7460(4) | 1 | 0.015(1)   | 8d |
| O202  | 0.3480(3) | 0.4952(6)  | 0.7434(4) | 1 | 0.016(1)   | 8d |
| O20   | 0.494(1)  | 0.394(3)   | 1.063(2)  | 1 | 0.15(1)    | 8d |
| O192  | 0.3334(3) | 0.2467(6)  | 0.4944(4) | 1 | 0.0150(1)  | 8d |
| O19   | 0.5       | -0.421(2)  | 0.75      | 1 | 0.051(5)   | 4c |
| O118  | 0.3336(3) | 0.366q(6)  | 0.3749(4) | 1 | 0.015(1)   | 8d |
| O182  | 0.5442(4) | 0.3541(4)  | 0.7102(5) | 1 | 0.022(2)   | 8d |
| O18   | 0.433(1)  | 0.551(4)   | 0.985(2)  | 1 | 0.15(1)    | 8d |
| O171  | 0.2530(3) | 0.3856(6)  | 0.7021(3) | 1 | 0.0115(9)  | 8d |
| O172  | 0.5471(4) | 0.2410(8)  | 0.8248(5) | 1 | 0.026(2)   | 8d |
| O17   | 0.4998(7) | 0.162(1)   | 0.4820(8) | 1 | 0.054(4)   | 8d |
| O162  | 0.3329(3) | 0.1261(6)  | 0.3756(4) | 1 | 0.013(1)   | 8d |
| O16   | 0.5736(4) | 0.004(2)   | 0.5170(6) | 1 | 0.052(4)   | 8d |
| O15   | 0.489(1)  | -0.272(2)  | 0.846(1)  | 1 | 0.099(8)   | 8d |

**Table II.** Ir-O bond distance inside the face-sharing dimers.

| Dimer #1 (consists of Ir01 and Ir03) | | | | | |
|---|---|---|---|---|---|
| Ir01-O13 (Å) | Ir01-O12 (Å) | Ir01-O151 (Å) | Ir01-O111 (Å) Shared with Ir03 | Ir01-O7 (Å) Shared with Ir03 | Ir01-O8 (Å) Shared with Ir03 |
| 1.900(7) | 1.885(7) | 1.900(7) | 2.062(7) | 2.071(7) | 2.060(7) |
| Ir03-O192 (Å) | Ir03-O118 (Å) | Ir03-O162 (Å) | Ir03-O111 (Å) Shared with Ir01 | Ir03-O7 (Å) Shared with Ir01 | Ir03-O8 (Å) Shared with Ir01 |
| 1.905(7) | 1.897(7) | 1.882(7) | 2.055(7) | 2.037(7) | 2.070(7) |
| Dimer #2 (consists of Ir02 and Ir04) | | | | | |
| Ir02-O10 (Å) | Ir02-O131 (Å) | Ir02-O112 (Å) | Ir02-O11 (Å) Shared with Ir04 | Ir02-O171 (Å) Shared with Ir04 | Ir02-O6 (Å) Shared with Ir04 |
| 1.905(7) | 1.927(7) | 1.916(7) | 2.051(7) | 2.070(7) | 2.069(7) |
| Ir04-O202 (Å) | Ir04-O141 (Å) | Ir04-O142 (Å) | Ir04-O11 (Å) Shared with Ir01 | Ir04-O171 (Å) Shared with Ir01 | Ir04-O6 (Å) Shared with Ir01 |
| 1.915(7) | 1.928(7) | 1.922(8) | 2.092(7) | 2.062(7) | 2.065(7) |

**Table III.** Ir-O bond distance of four single IrO$_6$ octahedrons.

| Octahedron #5 (consists of Ir05) | | | |
|---|---|---|---|
| Ir05-O17 (Å) | Ir05-O146 (Å) | Ir05-O16 (Å) | |
| 1.94(1) | 1.88(1) | 1.91(1) | |
| Octahedron #6 (consists of Ir06) | | | |
| Ir06-O152 (Å) | Ir06-O172 (Å) | Ir06-O182 (Å) | |
| 1.948(9) | 1.91(9) | 1.921(8) | |
| Octahedron #7 (consists of Ir07) | | | |
| Ir07-O154 (Å) | Ir07-O15 (Å) | Ir07-O155 (Å) | Ir07-O19 (Å) |
| 1.94(2) | 2.01(3) | 1.80(4) | 1.89(2) |
| Octahedron #8 (consists of Ir08) | | | |
| Ir08-O153 (Å) | Ir08-O20 (Å) | Ir08-O18 (Å) | |
| 1.86(1) | 1.80(4) | 1.82(3) | |

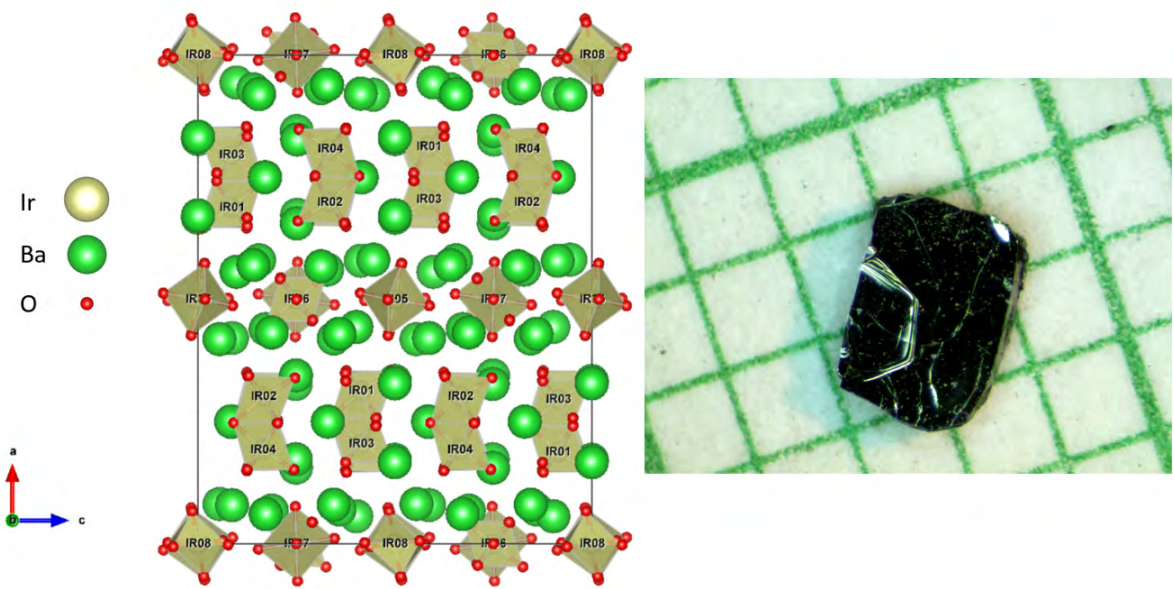

**Fig. 1**

**Captions:**

**Fig.1**. Left panel: the crystal structure of $Ba_{13}Ir_6O_{30}$. Right panel: a representative single crystal of $Ba_{13}Ir_6O_{30}$.